\documentclass[aps,prl,twocolumn]{revtex4}
\usepackage{amssymb}
\usepackage{graphicx}

\begin{document}

\title{Influence of  ambipolar potential on the properties of
inductively coupled discharges at the bounce resonance condition}

\author{Oleg V. Polomarov and Constantine E. Theodosiou}
\affiliation{Department of Physics and Astronomy, University of
Toledo, Toledo, Ohio, 43606-3390.}
\author{Igor D. Kaganovich }
\affiliation{Plasma Physics Laboratory, Princeton University,
Princeton, NJ 08543}

\date{\today }

\begin{abstract}

The importance of accounting for an ambipolar electrostatic
potential or a non-uniform density profile for modelling of
inductively coupled discharges is demonstrated.  A drastic
enhancement  is observed of the power transfer into plasma for
low-collisional, low-pressure, non-local discharges with
non-uniform electron density profiles under the condition of
bounce resonance.  This enhanced plasma heating is attributed to
the increase of the number of resonant electrons, for which the
bounce frequency inside the potential well is equal to the rf
field frequency.

\end{abstract}

\maketitle

\section{Introduction}

Low pressure radio-frequency (rf) inductive discharges have been
extensively used over the past decade as sources of inductively
coupled plasmas (ICP) in the plasma aided material processing
industry, semiconductor manufacturing, and lighting
\cite{Lieberman book, F.F.Chen book}. For very low pressures, i.e.
in the milliTorr region the ICP discharges exhibit a strong
non-local behavior and a number of peculiar physical effects
typical for warm plasmas, like  an anomalous skin penetration and
a resonant wave-particle interaction \cite{Lieber & Godyak review,
Kolobov review}.  The study of these effects  leads to further
optimization of the ICP sources and can result in improvement of
the characteristics of plasma-based devices.

An interesting effect that can lead to enhanced heating for
bounded, low-pressure plasmas is the possible bounce resonance
between the frequency $\omega$ of the driving rf field and the
frequency of the bounce motion of the plasma electrons confined in
the potential well by an ambipolar potential $\phi(x)$ and the
sheath electric fields near the discharge walls \cite{me APL,
Aliev and me, Ulrich and me, Me PRL 1998, Shaing, Chin WOOK, Chin
Wook EDF influence of bounce}. Most earlier theoretical and
numerical studies on this subject assumed for simplicity a uniform
plasma density over the discharge length, and the absence of an
ambipolar potential, i.e. allowed the electrons to bounce inside a
potential that is flat inside the plasma and infinite at the walls
\cite{Shaing, Godyak EDF, Chin WOOK, Chin Wook EDF influence of
bounce, Smolyakov}. Although these suppositions can result in a
fairly good description of the plasma behavior under non-resonant
conditions, the discharge parameters under resonant conditions can
be greatly altered by accounting for the presence of the ambipolar
potential, which -- it should be stressed -- always exists in real
discharges.  It is a very well known result of the quasilinear
theory, that for low-collisional discharges the plasma heating
essentially depends on the so-called ``resonant electrons," or
electrons with velocities equal to the phase velocities $v\simeq
\omega/k$ of the plane waves constituting the rf field (Landau
damping) \cite{Aliev and me,Ulrich and me}.  For bounded plasmas
and for an electron to be resonant, the above condition transforms
into the requirement that the rf field frequency must be equal to,
or be an integer multiple of the bounce electron frequency
$\omega=n \Omega_{b}$.  But the electron bounce frequency is very
sensitive to the actual shape of the ambipolar electrostatic
potential $\phi(x)$, especially for low-energy electrons.
Accounting for the electrostatic potential can lead the plasma
electrons into the resonant region even if they were not there in
the absence of the potential. This can result into a drastic
enhancement of the plasma heating and other related phenomena
\cite{Cluggish}.

In this article we present the results of a full, self-consistent
numerical modelling of low pressure ICP discharges under the
bounce resonance condition and show the pronounced influence of
the presence of the electrostatic ambipolar potential on the
plasma parameters under resonant conditions.

\section{Basic equations}

The model assumes a one-dimensional, slab geometry, inductively
coupled discharge of a plasma bounded on both sides by parallel
walls with a gap length $L$.  The walls carry fixed currents
flowing in opposite directions, produced by an external radio
frequency source.  The induced solenoidal rf electric field
$E_{y}$ is directed along the walls and the static ambipolar
electric field $E_{x}=-d\phi/dx$ of the ambipolar potential
$\phi(x)$ is directed towards the discharge walls, keeping
electrons confined and the plasma quasineutral, i.e.
$n_{e}(x)=n_{i}(x)$. In the present treatment of high density
discharge plasmas ($n_{e}\sim 10^{8} - 10^{12}~\textmd{cm}^{-3}$)
the sheath width is neglected, because it is of the order of a few
hundreds of microns, much less than the discharge dimension $L$.
Furthermore, it is assumed that the plasma electrons experience
specular reflection: a) from the discharge walls when they have
total energy $\varepsilon=mv^{2}/2-e \phi(x)$ larger than the
electron potential energy $-e\phi (x_{w})$ at the walls, $\pm
x_w$, and b) from the geometrical location of the turning points
$x_{\pm}(\varepsilon)$, where $-e\phi(x_{\pm})=\varepsilon$. The
above 1-D scheme can also be a good approximation for a
cylindrical ICP discharge, if the rf field penetration depth
$\delta$ into the plasma is less than the plasma cylinder radius
$R$ \cite{Meierovich}.

In order to describe the discharge self-consistently, one needs to
determine  the  rf electric  field profile $E_{y}(x)$, the
electron energy distribution function (EEDF) $f_{0}(\varepsilon)$,
and the ambipolar potential $\phi(x)$. The detailed description of
all the needed formalism is given in \cite{Our article}.  A short
account of the formalism is given below.

\subsection{Calculation of the EEDF }

For low-pressure discharges,  where the energy relaxation length
is large compared with the plasma width and the energy relaxation
time is large compared with the rf period,  the electron velocity
distribution function (EVDF) can be represented as a sum of the
main isotropic part $f=f_{0}(\varepsilon)$ (EEDF) that is a
function of only the total energy $\varepsilon$  and of a small
alternating anisotropic part $f_{1}(x,\mathbf{v},t)$,
$f=f_{0}(\varepsilon)+f_{1}(x,\mathbf{v},t)$ \cite{Tsendin 77 dc,
Me and Tsendin 1992 1, Me and Tsendin 1992 2}.  The Boltzmann
equation for the electron velocity distribution function reads
\begin{eqnarray}
\frac{\partial f_{1}}{\partial t}+ v_{x}\frac{\partial
f_{1}}{\partial x}+ \frac{e}{m}\frac{d\phi}{dx}\frac{\partial
f_{1}}{\partial v_{x}}- \frac{eE_{y}(x,t)}{ m}\frac{\partial
(f_{0}+f_{1})}{\partial v_{y}}\label{Vlasov eq.0}\\\nonumber=
St(f_{1}+f_{0}),
\end{eqnarray}
where $E_{y}(x,t)$ is the nonstationary rf electric field, and
$St(f)$ is the collision integral.  After applying the standard
quasilinear theory, Eq.(\ref{Vlasov eq.0}) splits into two
equations \cite{Aliev and me}, a linear one for $f_{1}$
\begin{equation}
\frac{\partial f_{1}}{\partial t}+v_{x}\frac{\partial
f_{1}}{\partial x} + \frac{e}{m}\frac{d\phi}{dx}\frac{\partial
f_{1}}{\partial v_{x}}-\frac{eE_{y}(x,t)}{m}\frac{\partial
f_{0}}{\partial v_{y}}=St(f_{1}),  \label{Eq. for f1}
\end{equation}
and a quasilinear one for $f_{0}$
\begin{equation}
-\overline{\frac{eE_{y}(x,t)}{m}\frac{df_{1}}{dv_{y}}}=\overline{St(f_{0})}.
\label{Eq. for f0}
\end{equation}
Here, the bar denotes space-time averaging over the phase space
available to electrons with total energy $\varepsilon $
\cite{Tsendin 77 dc, Me and Tsendin 1992 1, Me and Tsendin 1992
2}.  We can represent as  harmonic functions the rf electric field
$E_{y}(x,t)=E_{y0}(x)\exp (-i\omega t)$ and the anisotropic part
of the EVDF $f_{1}=f_{10}\exp (-i\omega t)$, where $\omega $ is
the discharge frequency.  Using the Bhatnagar-Gross-Krook (BGK)
approximation \cite{Our article}, $St(f_{1})=-\nu f_{1}$, and
omitting the subscript $0$ in the amplitudes, Eq.~(\ref{Eq. for
f1}) can be rewritten as
\begin{equation}
-i\omega f_{1}+v_{x}\frac{\partial f_{1}}{\partial
x}|_{\varepsilon _{x}}-ev_{y}E_{y}(x)\frac{df_{0}}{d\varepsilon
}=-\nu f_{1}, \label{Vlasov 1}
\end{equation}
where $\nu $ is the transport collision frequency,  $\varepsilon
_{x}=mv_{x}^{2}/2+\varphi(x)$ is the total energy along the
$x$-axis, and $\varphi(x)=-e\phi(x)$ is the electron potential
energy.

Eq.~(\ref{Vlasov 1}) can be effectively solved using a Fourier
series expansion.  Introducing the variable angle of the bounce
motion \cite{Me and Tsendin 1992 2} which is proportional to the
time of flight of an electron from the left turning point to the
current point
\begin{equation}
\theta (x,\varepsilon _{x})=\frac{\pi
\textrm{sgn}(v_{x})}{T(\varepsilon _{x})}
\int_{x_{-}}^{x}\frac{dx}{\left\vert v_{x}(\varepsilon
_{x})\right\vert }, \label{angle}
\end{equation}
where $T$ is half of the bounce period of the electron motion in
the potential well $\varphi (x)$
\begin{equation}
T_b(\varepsilon _{x})=\int_{x_{-}}^{x_{+}}\frac{dx}{\left\vert
v_{x}(\varepsilon _{x})\right\vert },  \label{bounce period}
\end{equation}
Eq.~(\ref{Vlasov 1}) simplifies to
\begin{equation}
-i\omega f_{1}+\Omega _{b}\frac{\partial f_{1}}{\partial \theta }
|_{\varepsilon _{x}}-v_{y}eE_{y}(\theta
)\frac{df_{0}}{d\varepsilon }=-\nu f_{1}.  \label{Vlasov 2}
\end{equation}
where $\Omega _{b}(\varepsilon _{x})=\pi /T(\varepsilon _{x})$ is
the bounce frequency for the electron in the potential well.
Making use of the Fourier series
\begin{equation}
g(x,\varepsilon _{x})=\sum_{n=-\infty }^{\infty }g_{n}\exp \left(
in\theta \right) ,  \label{Fourier direct}
\end{equation}
\begin{equation}
g_{n}=\frac{1}{2\pi }\left[ \int_{-\pi }^{\pi }g(\theta
,\varepsilon _{x})\exp \left( -in\pi \theta \right) d\theta
\right], \label{Fourier back}
\end{equation}
Eq.~(\ref{Vlasov 2}) gives
\begin{equation}
E_{yn}(\varepsilon _{x})=\frac{1}{\pi }\left[ \int_{0}^{\pi
}E_{y}(\theta )\cos \left( n\theta \right) d\theta \right] ,
\label{En}
\end{equation}
and
\begin{eqnarray}
f_{1s}(x,\varepsilon
_{x})&\equiv&1/2(f_{1}(v_{x}>0)+f_{1}(v_{x}<0))\\\nonumber
&=&-mv_{y}V_{y}^{\textmd{rf}}(x,\varepsilon _{x})\frac{df_{0}
}{d\varepsilon },
\end{eqnarray}
where
\begin{equation}
V_{y}^{\textmd{rf}}(x,\varepsilon
_{x})=-\frac{e}{m}\sum_{n=-\infty }^{\infty }\frac{ E_{yn}\cos
[n\theta (x)]}{in\Omega _{b}-i\omega +\nu }. \label{f1s Fourier}
\end{equation}

Knowing the symmetrical part $f_{1s}$ of the anisotropic
contribution to the EVDF, one can average Eq.~(\ref{Eq. for f0})
according to
\begin{eqnarray}
\overline{\textmd{Term}(x,\mathbf{v})}(\varepsilon )
&=&\int_{x_{-}}^{x_{+}}dxv(x,\varepsilon
)\textmd{Term}[x,v(x,\varepsilon )],
\label{averaging v} \\
v(x,\varepsilon ) &=&\sqrt{2[\varepsilon -\varphi (x)]/m}.
\end{eqnarray}
and obtain the final equation for $f_{0}$
\begin{eqnarray}
-\frac{d}{d\varepsilon }\left( D_{\varepsilon
}+\overline{D_{ee}}\right) \frac{df_{0}}{d\varepsilon
}-\frac{d}{d\varepsilon }\left[ \overline{V_{ee}}+
\overline{V_{el}}\right] f_{0}=\\\nonumber
 \sum_{k}\left[
\overline{\nu _{k}^{\ast }(w+\varepsilon _{k}^{\ast
})\frac{\sqrt{(w+\varepsilon _{k}^{\ast })}}{
\sqrt{w}}}f_{0}(\varepsilon +\varepsilon _{k}^{\ast
})-\overline{\nu _{k}^{\ast }}f_{0}\right]. \label{final averaged
k.e.}
\end{eqnarray}
Here, the bar denotes averaging according to Eq.~(\ref{averaging
v}), and $\nu^{*}_{k}$ is the inelastic collision frequency.  The
coefficients $V_{el},D_{ee},V_{ee}$ stem from the elastic and
electron-electron collision integrals, respectively, and are given
by \cite{Me and Tsendin 1992 1, Gurevich}
\begin{equation}
V_{el}=\frac{2m}{M}w\nu ,  \label{vel}
\end{equation}
\begin{equation}
V_{ee}=\frac{2w\nu _{ee}}{n}\left( \int_{0}^{w}dw\sqrt{w}f\right)
, \label{Vee}
\end{equation}
\begin{equation}
D_{ee}=\frac{4}{3}\frac{w\nu _{ee}}{n}\left(
\int_{0}^{w}dww^{3/2}f+w^{3/2}\int_{w}^{\infty }dwf \right),
\label{Dee}
\end{equation}
\begin{equation}
\nu _{ee}=\frac{4\pi \Lambda _{ee}n}{m^{2}v^{3}},  \label{nu ee}
\end{equation}
where $w=mv^{2}/2$ is the electron kinetic energy, $\nu _{ee}$ is
the Coulomb collision frequency, and $\Lambda _{ee}$ is the
Coulomb logarithm.

The energy diffusion coefficient responsible for the electron
heating is given by
\begin{eqnarray}
D_{\varepsilon }(\varepsilon )=\frac{\pi
e^{2}}{4m^{2}}\sum_{n=-\infty }^{\infty }\int_{0}^{\varepsilon
}d\varepsilon _{x} \\\nonumber \times\left\vert E_{yn}(\varepsilon
_{x})\right\vert ^{2}\frac{\varepsilon -\varepsilon _{x}}{\Omega
_{b}(\varepsilon _{x})}\frac{\nu }{\left[ \Omega _{b}(\varepsilon
_{x})n-\omega \right] ^{2}+\nu ^{2}}. \label{energy diffusion
coefficient nu constant}
\end{eqnarray}%
Note that this expression for $D_{\varepsilon }(\varepsilon )$
accounts for the bounce resonance $\Omega _{b}(\varepsilon
_{x})n=\omega $. The dependance of electron plasma heating on
resonant electrons especially pronounced for the $\nu<<\omega$, as
in this case
\begin{equation}
\frac{\nu }{\left[ \Omega _{b}(\varepsilon _{x})n-\omega \right]
^{2}+\nu ^{2}}\rightarrow\pi \delta( \Omega _{b}n-\omega )
\label{delta function}
\end{equation}%
where $\delta()$ is a Dirac delta function. It is worth to note
that if $L\rightarrow\infty$, the summation in (\ref{energy
diffusion coefficient nu constant})  goes into integration over
corresponding wave vectors $k_{n}$, and the bounce resonance
condition $\Omega _{b}(\varepsilon _{x})n=\omega $ transforms into
the well-known wave-particle  resonance condition for continuous
wave spectrum $k v=\omega$.

\subsection{Calculation of the rf electric field }
The transverse rf electric field $E_{y}$ is obtained from a single
scalar equation
\begin{equation}
\frac{d^{2}E_{y}}{dx^{2}}+\frac{\omega
^{2}}{c^{2}}E_{y}=-\frac{4\pi i\omega }{c^{2}}\left[ j(x)+I\delta
(x)-I\delta (x-L)\right] , \label{Maxwell eqs}
\end{equation}
where $I$ is the wall current and $j(x)$ is the induced electron
plasma current density that can be calculated knowing the
anisotropic part $f_{1s}$ of the EVDF
\begin{equation}
j=-\frac{em^{3/2}}{4\pi \sqrt{2}} \int f_{1s}v_{y}d^{3}\mathbf{v}.
\label{J definiton}
\end{equation}
Note that the normalization factor in Eq.(\ref{J definiton})
appears due to the normalization of $f_0$ as
\begin{equation}
n_{e}(x)=\int_{\varphi (x)}^{\infty }f_{0}(\varepsilon
)\sqrt{\varepsilon -\varphi (x)}d\varepsilon.  \label{n definiton}
\end{equation}
We now use the Fourier series
\begin{equation}
E_{y}(x)=\sum_{s=0}^{\infty }\Xi _{s}\cos (k_{s}x), \label{Fourier
series E}
\end{equation}%
where $s$ is an integer, $k_{s}=(2s+1)\pi /L$.  Substituting
Eq.~(\ref{Fourier series E}) into Eq.~(\ref{Maxwell eqs}) yields
\begin{equation}
\left( -k_{s}^{2}+\frac{\omega ^{2}}{c^{2}}\right) \Xi _{s}=-\frac{4\pi
i\omega }{c^{2}}\left[ j_{s}+\frac{2I\ }{L}%
\right] ,  \label{Eq. for Furier E}
\end{equation}
\begin{equation}
j_{s}=\frac{e^{2}}{m}\frac{n_e}{i(2s+1)\Omega
_{bT}}\sum_{l=0}^{\infty }\Xi _{l}Z_{s,l}^{\textmd{gen}}\left(
\frac{\omega +i\nu }{(2s+1)\Omega _{bT}}\right) , \label{J as
double summ}
\end{equation}
where $n_e$ is the plasma density at the discharge center, $\Omega
_{bT}=V_{T}\pi /L$, $V_{T}=\sqrt{2T/m}$,  and we introduced the
generalized plasma dielectric function \cite{Our article}
\begin{eqnarray}
Z_{s,l}^{\text{gen}}(\xi )&\equiv&
\sqrt{\frac{2}{m}}\frac{(2s+1)\pi \Omega _{bT}}{n_e L}
\sum_{n=-\infty }^{\infty }\int_{0}^{\infty }
\\\nonumber&\times&
 \frac{\Gamma (\varepsilon )}{
n\Omega _{b}(\varepsilon )-(2s+1)\Omega _{bT}\xi
}\frac{G_{s,n}(\varepsilon )G_{l,n}(\varepsilon )}{\Omega
_{b}(\varepsilon )}d\varepsilon,
\end{eqnarray}
where
\begin{equation}
\Gamma (\varepsilon )=\int_{\varepsilon}^{\infty
}f_{0}(\varepsilon )d\varepsilon. \label{gamma}
\end{equation}
In the limit of a uniform plasma, the generalized dielectric
function coincides with the standard plasma dielectric function
\cite{Our article}. The coefficients $G_{l,n}(\varepsilon )$ are
the temporal Fourier transforms of $\cos (k_{l}x)$ in the bounce
motion of an electron in the potential well ($m d^2x/dt^2=e d\phi/dx$)%
\begin{equation}
G_{l,n}(\varepsilon )=\frac{1}{T}\left[ \int_{0}^{T}\cos
[k_{l}x(\tau )]\cos \left( \frac{\pi n\tau }{T}\right) d\tau
\right].  \label{G coefficients}
\end{equation}
The Maxwell equation (\ref{Eq. for Furier E}) together with the
equation for the electron current (\ref{J as double summ}) and
(\ref{G coefficients}) comprise the complete system for
determining the profiles of the rf electric field.

\subsection{Calculation of the electrostatic potential}
 The electrostatic potential is obtained using the
quasineutrality condition
\begin{equation}
n_{e}(x)=n_{i}(x)=\int_{\varphi (x)}^{\infty }f_{0}(\varepsilon
)\sqrt{\varepsilon -\varphi (x)}d\varepsilon ,
\label{quasineutrality}
\end{equation}%
where $n_{e}(x)$ is the electron density profile and $n_{i}(x)$ is
the ion density profile given by a set of fluid conservation
equations for ion density and ion momentum \cite{badri and me}
\begin{equation}
\frac{\partial n_{i}}{\partial t}+\frac{\partial ( n_{i}
u_{i})}{\partial x}=R,\label{n_cont}
\end{equation}%
and
\begin{equation}
\frac{\partial (n_{i} u_{i})}{\partial t}+\frac{\partial ( n_{i}
u_{i} u_{i})}{\partial x}=-\frac{n_{i}}{M_{i}}\frac{\partial
\phi(x) }{\partial x}-\nu_{i} n_{i} u_{i}
 ,\label{u_moment}
\end{equation}%
where $R$ is the ionization rate, $\nu_{i}$ is the ion-neutral
collision frequency and $n_{i}, u_{i},$ and $M_{i}$ are ion
density, velocity, and mass, respectively.

Eq.~(\ref{quasineutrality}) is solved in the form of a
differential equation \cite{My CCP}
\begin{equation}
\frac{d\varphi }{dx}=-T_{e}^{\text{scr}}(x)\frac{d\ln
[n_{i}(x)]}{dx},\label{pot_diff}
\end{equation}%
where $T_{e}^{\text{scr}}(x)$ is the electron screening temperature%
\begin{equation}
T_{e}^{\text{scr}}(x)=\left[ \frac{1}{2n(x)}\int_{\varphi
(x)}^{\infty
}f_{0}(\varepsilon )\frac{d\varepsilon }{\sqrt{\varepsilon -\varphi (x)}}%
\right] ^{-1},
\end{equation}
and the electrostatic ambipolar potential can be obtained by
integration of Eq.(\ref{pot_diff}).

The above described self-consistent system of equations was
formulated in \cite{Our article}, and implemented and compared
with the experimental data in \cite{badri and me,badri and me 2}.
Although the simulation results of the latter articles were proven
to be adequate, the method of the direct computation of Green
functions, used there, seems to be impractical, because of the
excessively long computational time (about a day).  To speed up
the calculations (to about an hour), the Fast Fourier Transform
[Eqs.(\ref{Eq. for Furier E})-(\ref{G coefficients})] was used in
the present simulations.

\section{Results and discussion}

\begin{figure}
\includegraphics[width=3in]{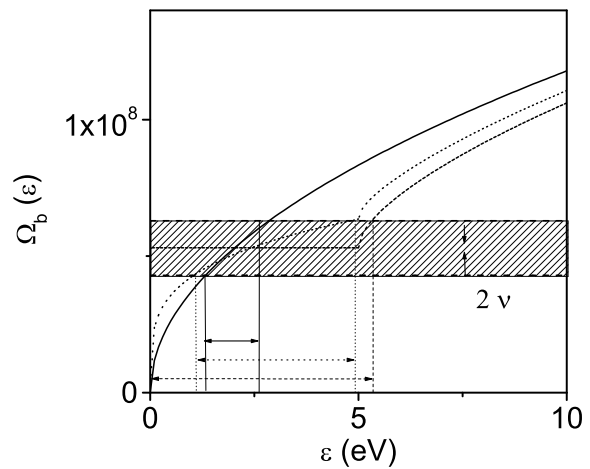}
\caption{The electron bounce frequency
$\Omega_{b}(\varepsilon_{x})=\pi/T_{b}(\varepsilon_{x})$, for
discharge of length $L=5~\text{cm}$, as a function of the electron
energy [$\varepsilon_{x}=mv_x^2/2-e\phi(x)$] for different
potential wells, consisting of the reflecting walls and different
ambipolar potentials $\phi(x)$. Solid line corresponds to a
uniform plasma with $\phi(x)=0$; dashed line -- quadratic
potential $\phi(x)=5\times (2 x/L-1)^{2}~\textmd{eV}$; and dotted
line -- quartic potential $\phi(x)=5\times (2
x/L-1)^{4}~\textmd{eV}$. Here, $T_{b}(\varepsilon_{x})=
\int_{x_{-}}^{x^{+}}dx/v_x,\varepsilon_{x}(\varepsilon_{x})$. The
cross-hatched box shows the resonance region.  Arrows show
electron energies in the resonance region.} \label{Fig.1}
\end{figure}

 Collisionless heating is a very important channel of
power transfer for bounded, warm, low-collisional plasmas
\cite{Lieber & Godyak review}. For a semi-infinite plasma, the
collisionless heating essentially depends on the resonant
electrons, i.e. electrons moving  with the velocities equal to
phase velocities of the components of the spectrum of the driving
rf field. Such electrons can effectively gain energy from a wave
and, henceforth, the plasma can be efficiently heated under
resonant conditions. For the case of  bounded plasma, the
condition for resonance heating transforms into the bounce
resonance condition, i.e., the frequency $\omega$ of the
electromagnetic wave must coincide, or to be several times larger
of the bounce frequency $\Omega_{b}$ of the electron bounce motion
in the potential well. If the electron mean free path is larger
than the discharge gap $L$, the resonant electrons with
$\Omega_{b}=\omega$ accumulate velocity changes in successive
interactions with the rf electric field, which lead to very
effective electron heating \cite{me APL}.

The importance of the resonant electrons for plasma heating under
the bounce resonance condition, can be readily examined from
expression (\ref{energy diffusion coefficient nu constant}) for
the electron energy diffusion coefficient, where the term $\Omega
_{b}(\varepsilon _{x})n=\omega $ in the denominator clearly shows
the dominant role of the resonant electrons for collisionless
electron heating. The total power $P$  deposited into plasma, per
unit square of a side surface, is related to the electron energy
diffusion coefficient $D_{\varepsilon }(\varepsilon )$ \cite{Our
article},
\begin{equation}
P=-\sqrt{2m}\int_{0}^{\infty }D_{\varepsilon }(\varepsilon )\frac{%
df_{0}(\varepsilon )}{d\varepsilon }d\varepsilon. \label{P(D)}
\end{equation}

The presence of an ambipolar  electrostatic potential can greatly
affect the electron heating due to two  reasons:

1) the ambipolar potential confines low energy electrons to the
center of the discharge plasma and these electrons cannot reach
the region of the strong field near the walls, and

2) the number of resonant electrons is generally larger for a
nonuniform plasma than for a uniform plasma due to influence of
the electrostatic ambipolar potential on the bounce frequency.

\begin{figure}
\includegraphics[width=3in]{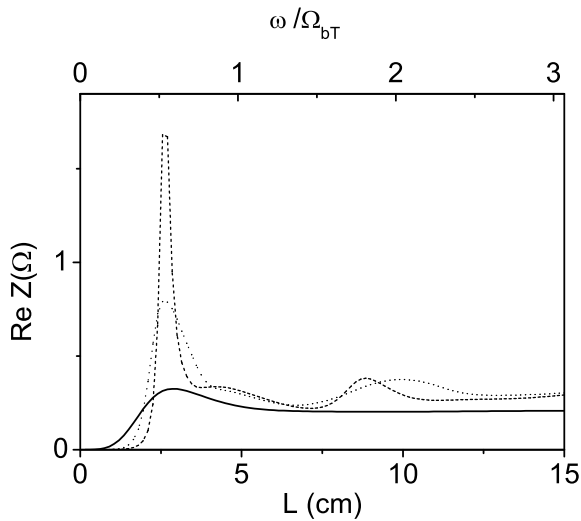}
\caption{The plasma resistivity   as a function of the distance
between the walls for a uniform plasma (without any ambipolar
potential) and a nonuniform plasma with the quadratic and quartic
potentials and a given Maxwellian EEDF. Discharge parameters are:
electron temperature $T_e=5~\textmd{eV}$, peak electron density at
the center of the discharge $n_{e}=5\times
10^{11}~\text{cm}^{-3}$, rf field frequency $\omega/2\pi=13.56~
\textmd{MHz}$, and electron transport frequency
$\nu=10^{7}~\textmd{s}^{-1}$.  The lines correspond to the same
cases as in Fig.~1.} \label{Fig.2}
\end{figure}
If the ambipolar potential is accounted for, then for low energy
electrons, the distance between  turning points is smaller than
the distance between walls $L$.  This results to an increase of
their bounce frequencies compared to the uniform plasma case.
Therefore, if electrons with low energy were far from the bounce
resonance in a uniform plasma, they can approach the resonance
region in nonuniform plasma. Fig.~\ref{Fig.1} shows the dependance
of the electron bounce frequency $\Omega_{b}(\varepsilon_{x})$ on
the electron energy $\varepsilon$ for different potential wells,
consisting of the reflecting walls and different ambipolar
potentials $\phi(x)$.  Here,
$\Omega_{b}(\varepsilon_{x})=2\pi/T_{b}(\varepsilon_{x})$, where
$T_{b}$ is half of the bounce period of the electron motion in the
potential well $\varphi (x)$ given by Eq.~(\ref{bounce period}).
The width of the resonance is given by Eq.(\ref{delta function})
and is proportional to $\nu$.  The population of resonant
electrons consists of all electrons corresponding to the interval
of bounce frequencies $\Omega _{b}(\varepsilon _{x})n \in
\left[\omega -\nu, \omega +\nu \right]$ where for most practical
cases $n=1$ -- only the first resonance is important. From
Fig.~\ref{Fig.1} it is apparent that the number of resonance
electrons increases if the ambipolar potential is accounted for.
For example, all electrons confined in the quadratic potential
have the same bounce frequency and are all resonant, if the bounce
frequency is equal to the discharge frequency.
\begin{figure}
\includegraphics[width=3in]{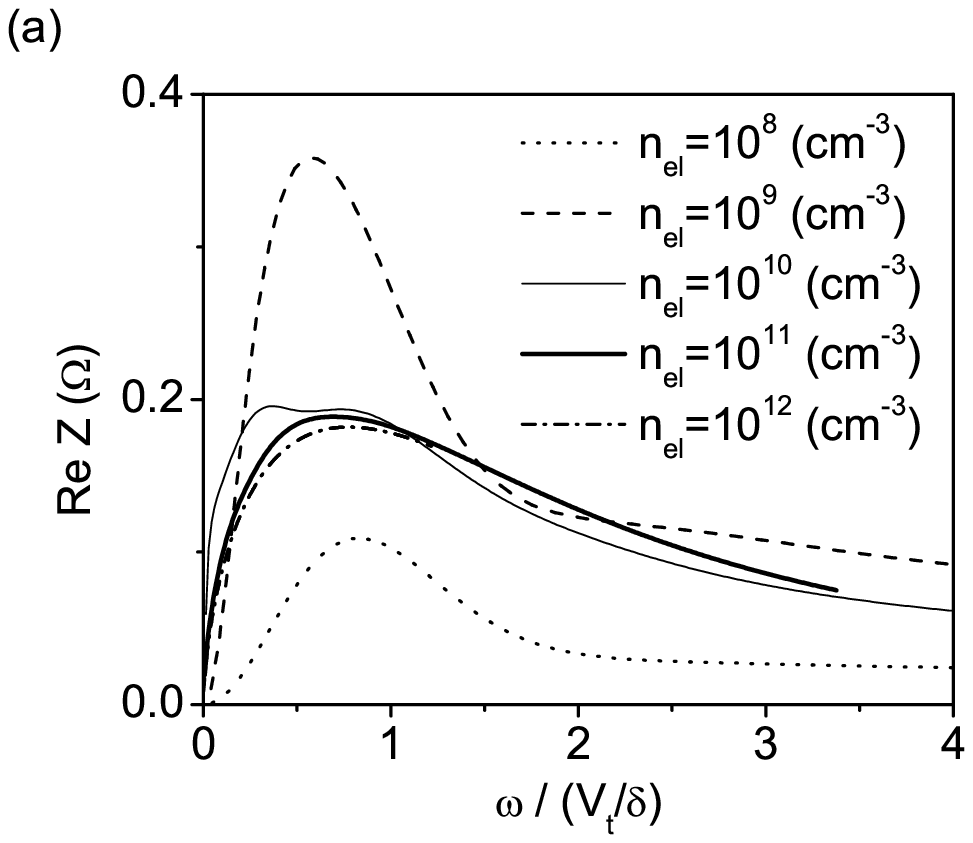}
\includegraphics[width=3in]{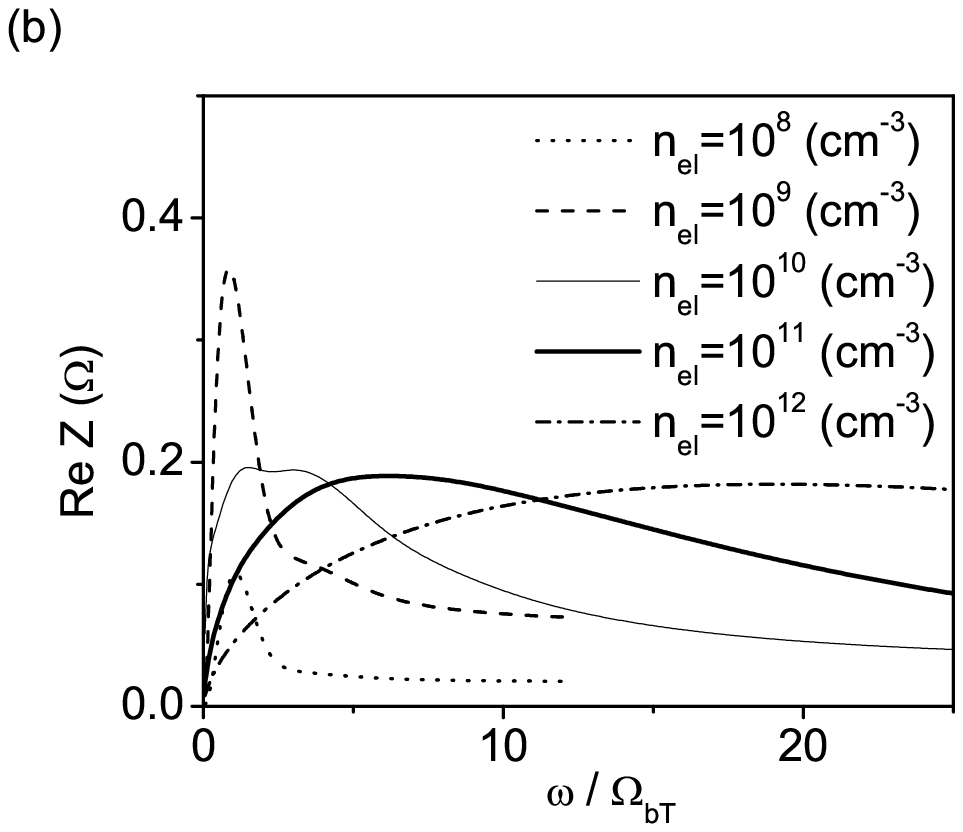}
\caption{The plasma resistivity  as a function of the driving
frequency normalized by (a) the inverse of the transit time
through the skin layer $V_{T}/\delta_{\text{max}}$ and (b) by the
bounce frequency $V_{T}\pi/L$, for different electron densities
and for the fixed discharge parameters: $L=50~\text{cm}$,
$T_{e}=5~ \textmd{eV}$ and $\nu=10^{6}~\textmd{s}^{-1}$. Here,
$V_{T}$ is the thermal velocity and $\delta_{\text{max}}$ is the
rf field plasma penetration depth for the maximal value of the
resistivity at a given density.}\label{Fig.3}
\end{figure}
The results of the calculations described below show that
increase of the number of the resonant electrons has a much more
profound effect on the discharge plasma heating than the mere
confinement of them in the region of low rf field.

\subsection{Non-self-consistent simulations with a given Maxwellian
EEDF and given ambipolar potentials} Most of earlier works on
low-pressure ICP discharges where reported assuming a Maxwellian
EEDF, uniform plasma  and accounted for the sheath electric field
by  bouncing electrons off the discharge walls \cite{Shaing,Godyak
EDF,Chin WOOK,Chin Wook EDF influence of bounce}.

In order to explicitly show the importance of accounting for
ambipolar potential on collisionless electron heating, we
performed numerical simulations using a given Maxwellian EEDF for
uniform and nonuniform plasmas (\textit{with} and \textit{without}
an ambipolar potential). Specifically, we obtained results for the
dependence on the plasma length of the plasma resistivity or the
real part of impedance $Z=4 \pi/c\times E(0)/B(0)$, where $E(0)$
and $B(0)$ are the electric and magnetic field at the wall,
respectively \cite{Kolobov review}. The surface impedance is
related to the power deposition according to:
\begin{equation}
 P=-I^2 \text{Re}Z \label{PonI},
\end{equation}
 where $I$ is the amplitude of the current.

The results are presented in Fig.~\ref{Fig.2}.  It is clearly seen
that the presence of the ambipolar potential enhances
significantly the resistivity of the plasma under the bounce
resonance condition, compared to the case of the absence of a
potential. The most profound change in resistivity is observed for
the quadratic potential.  In this latter case all trapped
electrons have the same bounce frequency, and thus all of them can
be resonant.  The obtained results explicitly show that neglecting
the ambipolar potential, as is often done for simplicity, can lead
to large discrepancies (more than 100 percent), especially for
conditions close to the bounce resonance. For large gaps $L \gg
2\pi\delta = 5~\text{cm} $ (where the skin depth is
$\delta=0.79~\text{cm}$ for the conditions in Fig.~\ref{Fig.2}),
the two skin layers near both walls are independent of each other,
as the gap width is much larger than the nonlocality length
$l=V_T/\omega=1.55~\text{cm}$. As a result, for $L>5~\text{cm}$
the surface impedance does not depend on the gap width. In the
opposite limit $L \ll 2\delta$, the electric field profile is
linear $E_y=E_y(0) (1-2x/L)$ and the plasma resistivity is mostly
determined by the first bounce resonance $\omega=\Omega_b$. For
uniform plasmas, only slow electrons contribute to the
collisionless heating, because the resonant velocity corresponds
to $v_x=\omega L/\pi=V_T L/\pi l<<V_T$. If $L \ll 2\delta$ the
number of resonant electrons and the total number of electrons in
plasma are decreasing for smaller $L$, which leads to a smaller
heating. For nonuniform plasmas, the bounce resonance condition
$\omega=\Omega_b$ can not be satisfied for any electron energy for
small $L<2~\text{cm}$ and collisionless heating vanishes, see
Fig.~\ref{Fig.2}.

The maximum of plasma resistivity occurs at $L \sim 2\delta$.
Similarly to the case of short gaps, the real part of the surface
impedance is mostly due to the first bounce resonance
$\omega=\Omega_b$, and only slow electrons contribute to
collisionless heating for uniform plasma, as $v_x \sim 2\delta
\omega /\pi=V_T 2\delta/\pi l<V_T$. For nonuniform plasmas, the
bounce frequency is higher than in uniform plasmas and most of
electrons are in resonance. As a result the surface resonance
plotted as a function of the gap width has a pronounced peak
compared to the shallow maximum in uniform plasmas, see
Fig.~\ref{Fig.2}.

To compare the condition of the bounce resonance $\omega=\Omega_b$
with the corresponding condition of the transit resonance
$\omega=V_{T}k$, additional simulations of the dependance of the
plasma resistivity Re$Z$ on the driving frequency $\omega$ were
performed for the fixed length $L=50~\textmd{cm}$ with given
Maxwellian EEDF corresponding to the electron temperature $T_e=5~
\textmd{eV}$, and various electron densities for a uniform plasma
(without accounting for the electrostatic potential). In
Fig.~\ref{Fig.3}(a) the resistivity is shown as a function of the
discharge frequency normalized by the inverse of the ``transit"
time of the electron pass through the skin layer
$V_{T}/\delta_{\text{max}}$, where $\delta_{\text{max}}$ is the
plasma penetration depth of the rf field for the maximal value of
the resistivity defined as
$\delta=\int_{0}^{L/2}dx|[E_{y}(x)/E_{y}(0)|$. In
Figure~\ref{Fig.3}(b) the plasma resistivity is plotted versus the
driving frequency normalized by the thermal bounce frequency
$\Omega_{bT}=V_T\pi/L$. At high plasma densities ($>10^{10}
\text{cm}^{-3}$) the field penetration depth is much smaller than
the discharge gap $L$ and the effects of the finite discharge gap
are unimportant: the maximum of the surface impedance and,
correspondingly, the most efficient collisionless electron heating
occurs at the ``transit" resonance $\omega\simeq V_{T}/\delta$, as
for the case of semi-infinite plasmas.  At low electron densities
(and frequencies) ($n_e<10^{10}~ cm^{-3}$) the electric field
penetration depth is of the order of the discharge dimension
$\delta \sim L$, and as a result the maximum of the surface
impedance corresponds to the condition of the bounce resonance
$\omega\simeq \pi V_{T}/L$. Note that the absolute maximum of the
power dissipation occurs when both conditions for the bounce and
transit resonances are met, which occurs for $L=\pi\delta$
\cite{Our article}. This happens at
$n_{e}\leq10^{9}~\text{cm}^{-3}$ for the conditions of
Fig.~\ref{Fig.3}.  At high frequencies $\omega \gg \pi
V_{T}/\delta$, the number of resonant electrons is exponentially
small and collisionless heating vanishes.  The resulting heating
depends on the collision frequency and nonlocal effects, as
described in Ref. \cite{me APL}.

\begin{figure}
\includegraphics[width=3in]{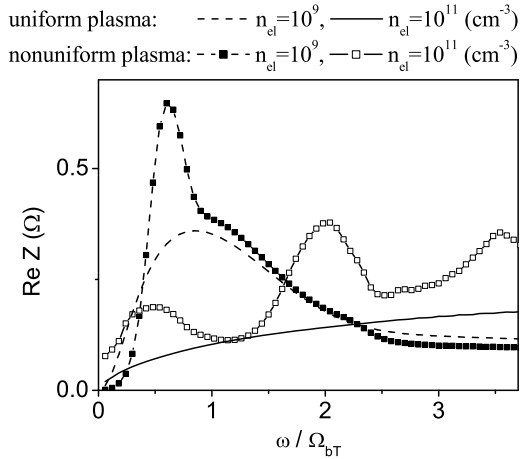}
\caption{The plasma resistivity  as a function of the driving
frequency normalized by the thermal bounce frequency for different
electron densities \textit{with} and \textit{without} the
ambipolar potential  $\phi(x)=5\times (2 x/L-1)^{4}~\textmd{eV}$
and for the same discharge parameters as in Fig.3.} \label{Fig.4}
\end{figure}

Accounting for the ambipolar potential enhances considerably the
plasma resistivity for the aforementioned case. As  shown in
Fig.~\ref{Fig.4}, the pronounced maxima of Re$Z$ appear for the
frequencies that correspond to  integer multiples of the bounce
frequency $\omega= n \Omega_{bT}$, because accounting for the
electrostatic potential yields a larger number of the resonant
electrons.
\subsection{Self-consistent calculations}
\begin{figure}
\includegraphics[width=3in]{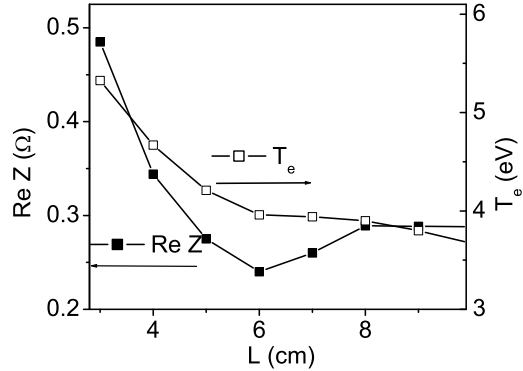}
\caption{The results of self-consistent simulations for the plasma
resistivity and the electron temperature (defined as 2/3 of the
average electron energy) at the center of the discharge as
functions of the size of the discharge gap and for a given surface
current $I=5~\text{A/m}$, rf field frequency $\omega/2\pi=13.56~
\textmd{MHz}$ and argon pressure $P=3~\textmd{mTorr}$.}
\label{Fig.5}
\end{figure}

\begin{figure}
\includegraphics[width=3in]{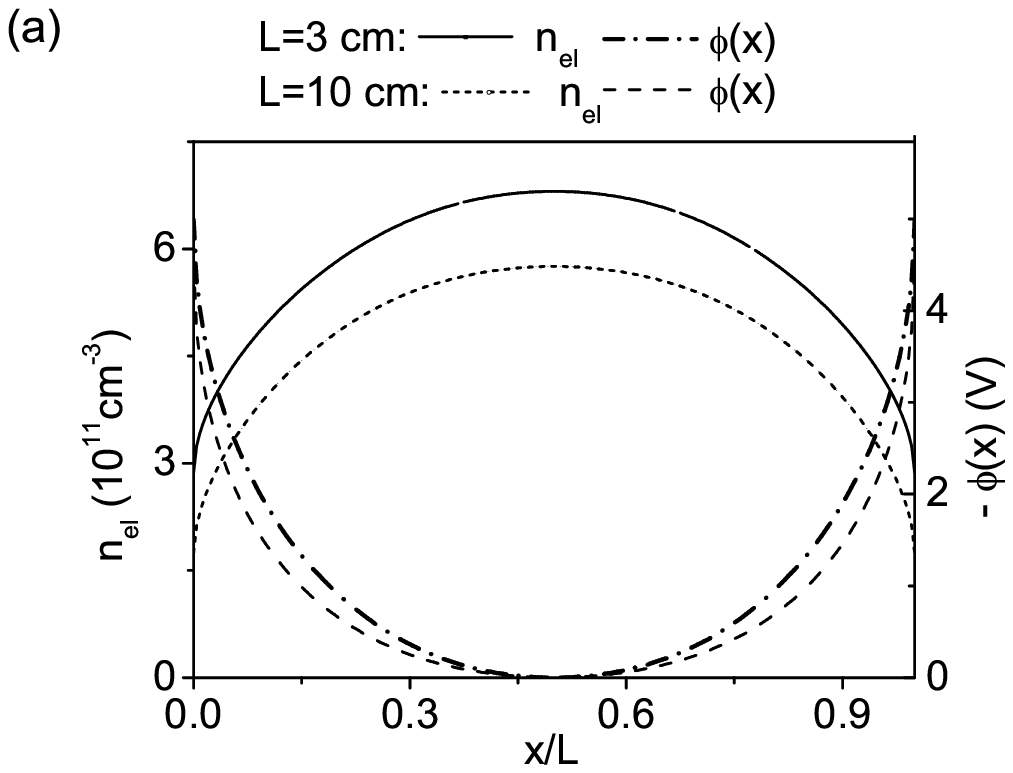}
\includegraphics[width=3in]{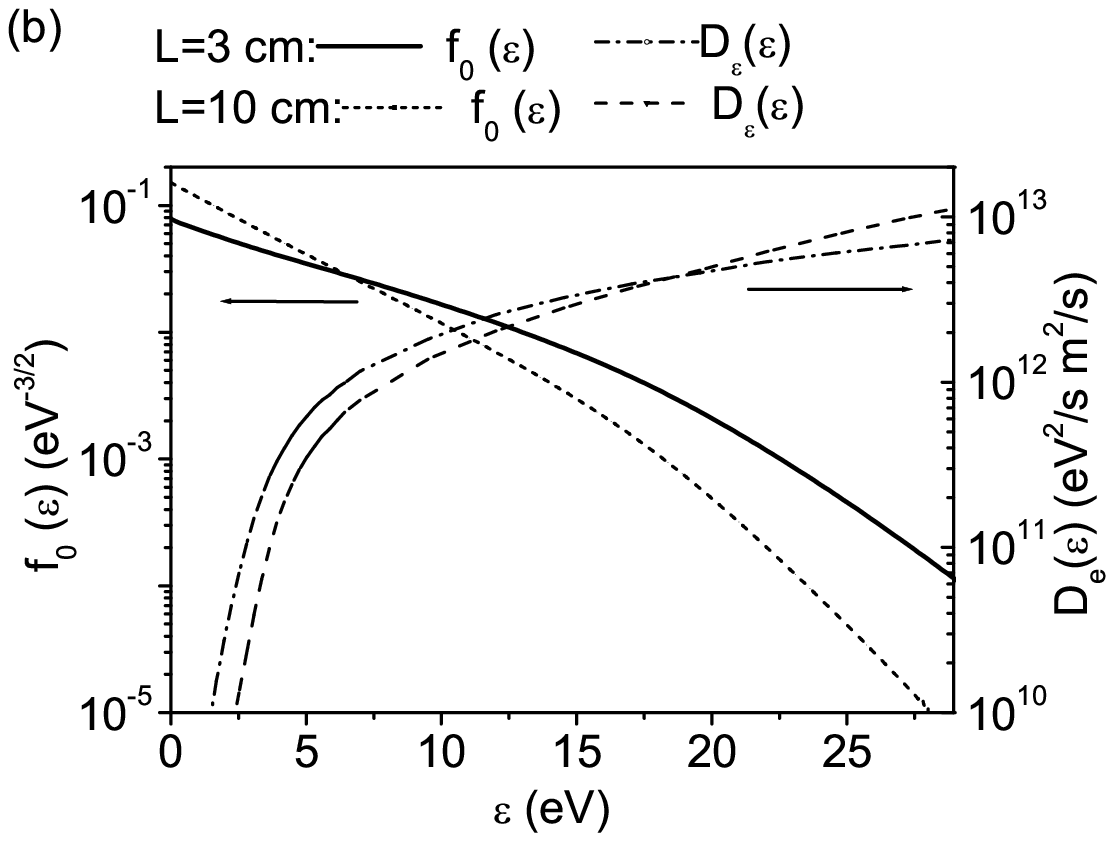}
\caption{The results of self-consistent simulations for the
discharge gap $L=3~\text{cm}$ corresponding to the bounce
resonance, and $L=10~\text{cm}$ for the same conditions as in
Fig.5 (a) the electron density and ambipolar potential profiles,
(b)the electron energy distribution function EEDF and the energy
diffusion coefficient $D_{\varepsilon}(\varepsilon)$ profiles. }
\label{Fig.6}
\end{figure}

To investigate the behavior of discharge parameters under the
condition of a bounce resonance, the full self-consistent
simulations of the EEDF,  rf electric field, and  ambipolar
potential for  fixed surface current have been performed for
$13.56~\text{MHz}$ rf driving frequency. Figure \ref{Fig.5} shows
the dependence of the resistivity of the discharge plasma on the
discharge dimension. The calculations have been performed for
discharge gaps from $3~\text{cm}$ to $10~\text{cm}$ (the discharge
can not be sustained for gaps smaller then $3~\text{cm}$ at a
given pressure of $3~ \text{mTorr}$). It can be clearly seen that
the resistivity of the plasma sharply increases for the discharge
gap corresponding to the bounce resonance (about $3~\text{cm}$).
The self-consistent electrostatic potential and ion-electron
density profiles are plotted in Fig.~\ref{Fig.6}(a) for two
different discharge lengths - $3~\text{cm}$, corresponding to the
bounce resonance condition, and $10~\text{cm}$, corresponding to
the non-resonant width. These graphs show that the electron
density at the center of the discharge is larger for the
$3~\text{cm}$ resonant gap than for the $10~\text{cm}$
non-resonant gap. Note that if the power transfer efficiency, or
the surface impedance, were the same, then the total power
transferred into the plasma would also be the same and the plasma
densities would be equal, due to energy balance.  In our case the
surface impedance for $3~\text{cm}$ gap is considerably higher,
what corresponds to the higher plasma density.

\begin{figure}
\includegraphics[width=3in]{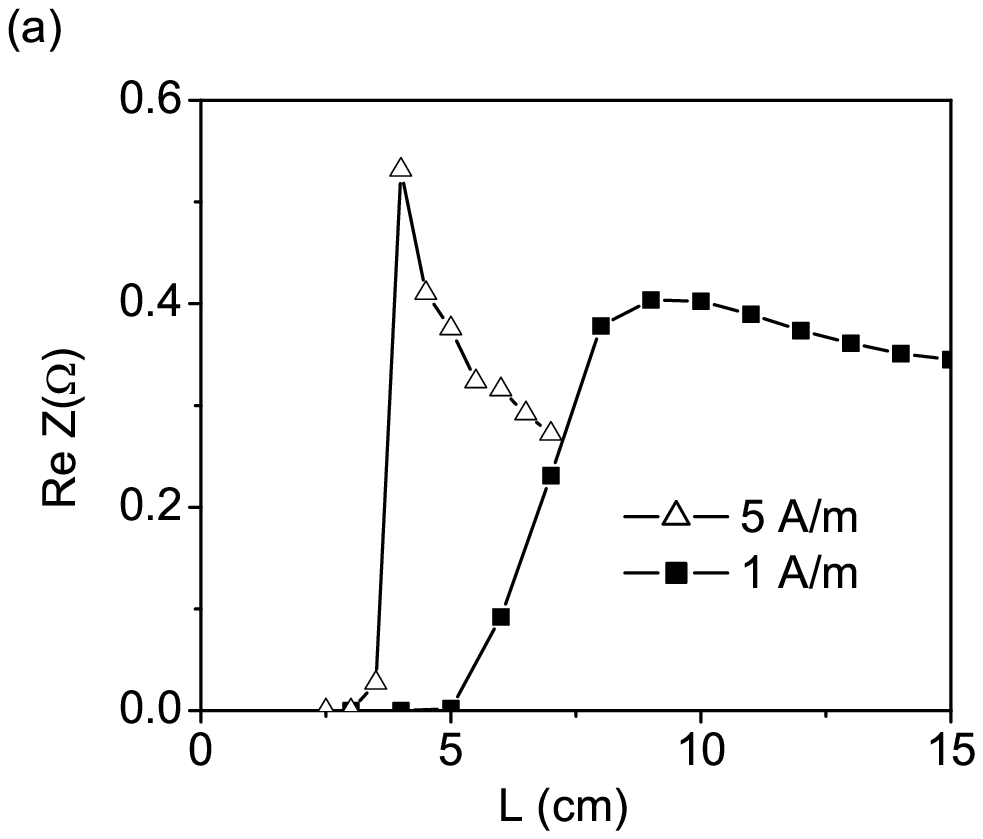}
\includegraphics[width=3in]{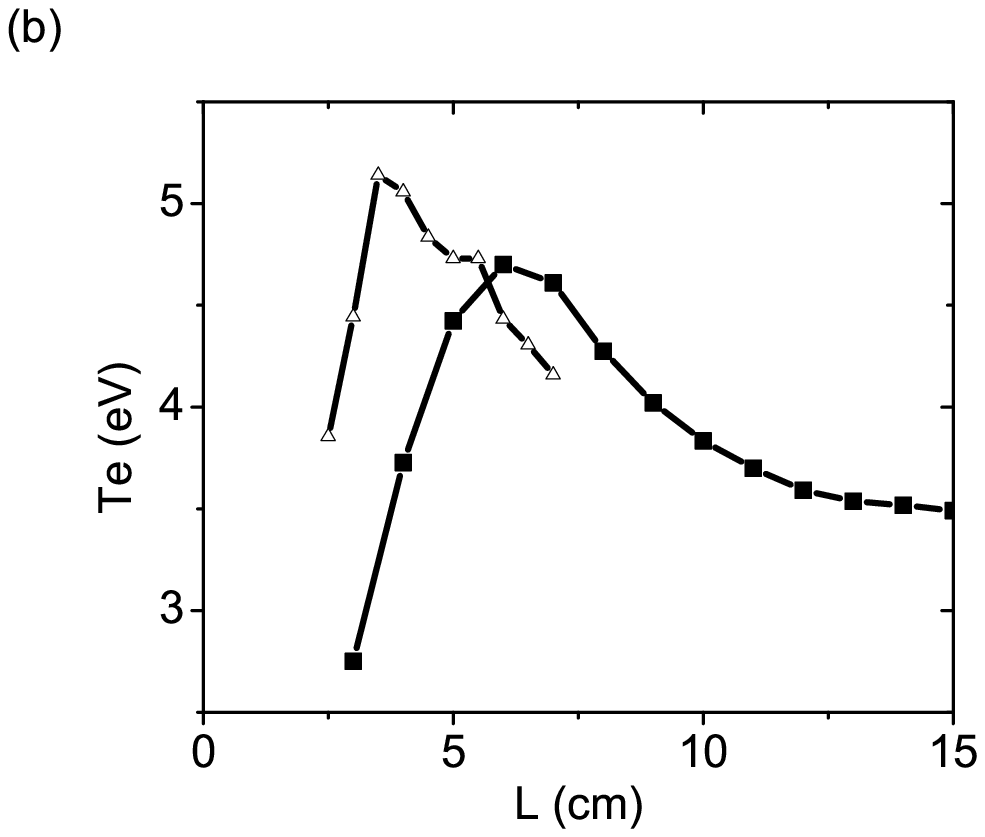}
\caption{The self-consistent simulations for two different surface
currents $I=1~ \textmd{A/cm}$ and $ I=5~ \textmd{A/cm}$ and the
given discharge parameters: $P=3~ \textmd{mTorr}$,
$\omega/2\pi=6.78~\textmd{MHz}$. Shown are (a) the plasma
resistivity  and (b) the electron temperature at the discharge
center (two thirds of average electron energy) as functions of the
discharge gap.} \label{Fig.7}
\end{figure}

\begin{figure}
\includegraphics[width=3in]{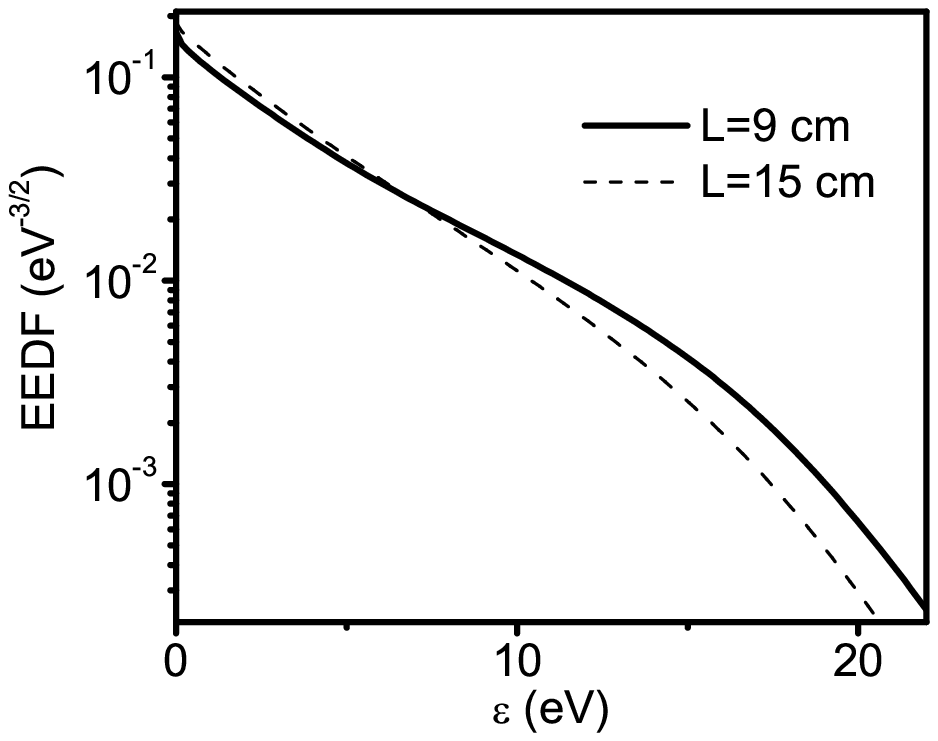}
\caption{The electron energy distribution functions  for the
bounce resonance discharge gap length, $L=9~\textmd{cm}$, and for
a nonresonant discharge gap $L=15~\text{cm}$, for the surface
current $I=1~\textmd{A/cm}$; the other conditions are the same as
in Fig.~7. } \label{Fig.8}
\end{figure}

The electron energy distribution function and the diffusion
coefficient in energy space are shown in Fig.~\ref{Fig.6}(b).
Figure \ref{Fig.6}(b) shows that the energy diffusion coefficient
is larger for the $3~\text{cm}$ gap than for the $10~\text{cm}$
gap for electron energy less than $15~\text{eV}$. This results in
more effective electron heating, leading to the larger plasma
resistivity shown in Fig.~\ref{Fig.5}.
The steady-state electron energy distribution function is governed
by the following processes: the collisionless electron heating in
the rf electric field, inelastic collisions with neutrals, and
redistribution of energy among plasma electrons due to
electron-electron collisions. We see in Fig.~\ref{Fig.6}(b) that
the  EEDF shape is similar to the two-temperature EEDF
\cite{Lieber & Godyak review} with the temperature of the tail of
the distribution being lower than the temperature of the main body
of the EEDF, corresponding to the onset of inelastic collisional
losses. For a $3~\text{cm}$ gap, corresponding to the bounce
resonance condition, the electron temperature of low-energy
electrons (less than the excitation potential $11.5~\text{eV}$) is
much higher than for the $10~\text{cm}$ non-resonant gap.  This
effect is similar to the plateau formation on the EEDF governed by
collisionless heating in the finite range of electron energies
\cite{Aliev and me}.  Under the conditions of Fig.~\ref{Fig.6}
this plateau is not well pronounced, because it is smeared out by
electron-electron collisions.
Additional simulations have been performed also for the discharge
frequency $6.78~\text{MHz}$, which is half the driving rf field
frequency considered above. For the lower driving frequency the
first bounce resonance shifts in the region of larger $L$, as
shown in Fig.~\ref{Fig.7}. Figure \ref{Fig.7} shows the calculated
resistivity for two different surface currents, $1~\textmd{A/cm}$
and $5~\textmd{A/cm}$.  One can see that the positions of the
resistivity maxima corresponding to different surface currents are
shifted relatively to each other. The larger surface current
corresponds to a larger power transfer into the plasma according
to Eq.(\ref{PonI}) and results in a higher plasma density ($n \sim
I^2$, $n_e=2\times 10^{10}$ and $n_e=7\times 10^{11}$,
respectively).  The higher discharge plasma density, in turn,
leads to a smaller skin depth.  Correspondingly, the position of
resistivity maximum shifts into the region of smaller discharge
gaps of the order of the skin depth.
The electron energy distribution functions for $6.78~\textmd{MHz}$
are plotted in Fig.~\ref{Fig.8} for the surface current
$1~\textmd{A/m}$ and for two different lengths, resonant
$9~\text{cm}$ and non-resonant $15~\text{cm}$. The phenomenon of
plateau-formation on the EEDF is clearly seen for the bounce
resonance condition for $L=9~\text{cm}$.
\section{Conclusion}
The analysis of the properties of inductively coupled discharges
clearly shows  the phenomenon of the bounce resonance.
Self-consistent simulations of the discharge resistivity and
electron energy distribution functions demonstrate the
significance of the explicit accounting for the non-uniform plasma
density profile and the correct form of the ambipolar
electrostatic potential. Enhanced electron heating and larger
plasma densities (for a given current in the coil) can be achieved
if the low-pressure ICP discharges are operated under the
conditions of the bounce resonance.

{}

\end{document}